\documentclass[a4paper,11pt]{article}
\pdfoutput=1 

\usepackage{jcappub} 

\usepackage[T1]{fontenc} 
\usepackage{latexsym}
\usepackage{amssymb}
\usepackage{epsfig}
\usepackage{amsmath}
\usepackage{float}
\usepackage{color}
\usepackage{enumitem}
\usepackage{placeins}
\usepackage{hhline}

\newcommand{\be}{\begin{equation}}
\newcommand{\ee}{\end{equation}}
\newcommand{\bea}{\begin{eqnarray}}
\newcommand{\eea}{\end{eqnarray}}

\definecolor{MLc}{rgb}{0.,.6,0.}
\newcommand{\cML}{\color{black}}

\newcommand{\Fb}{\bar{F}}
\newcommand{\Zb}{\bar{Z}}

\newcommand{\MontePython}{\textsc{MontePython}}
\newcommand{\CLASS}{\textsc{class}}
\newcommand{\Lcdm}{$\Lambda$CDM}

\begin{document}

\title{Variable sound speed in interacting dark energy models}

\author{Mark S. Linton,$^{a}$}
\author{Alkistis Pourtsidou,$^{b,a}$}
\author{Robert Crittenden,$^{a}$}
\author{Roy Maartens $^{c,a}$}
\affiliation{
$^a$ Institute of Cosmology and Gravitation, University of Portsmouth,\\
 Dennis Sciama Building, Burnaby Road, Portsmouth, PO1 3FX, United Kingdom \\
$^b$ School of Physics and Astronomy, Queen Mary University of London, \\Mile End Road, London E1 4NS, United Kingdom \\
$^c$ Department of Physics and Astronomy, University of the Western Cape,\\ Cape Town 7535, South Africa
}
\emailAdd{mark.linton@port.ac.uk}
\emailAdd{a.pourtsidou@qmul.ac.uk}
\emailAdd{robert.crittenden@port.ac.uk}
\emailAdd{roy.maartens@gmail.com}
\begin{abstract}{
We consider a self-consistent and physical approach to interacting dark energy models  described by a Lagrangian, and identify a new class of models with variable dark energy sound speed. We show that if the interaction between dark energy in the form of quintessence and cold dark matter is purely momentum exchange this generally leads to a dark energy sound speed that deviates from unity. Choosing a specific sub-case, we study its phenomenology by investigating the effects of the interaction on the cosmic microwave background and linear matter power spectrum. {\cML We also perform a global fitting of cosmological parameters using CMB data, and compare our findings to $\Lambda$CDM}.}
\end{abstract}
\maketitle

\section{\label{sec:Intro}Introduction}

Discovering the nature of dark energy is arguably one of the main goal of modern cosmology. Whether the observed accelerated expansion of the Universe is due to a cosmological constant, a dynamical field like quintessence (see \cite{Copeland:2006wr} and references therein), a signature of modified gravity (see \cite{CliftonEtal2011} and references therein), or some more exotic or undiscovered phenomena, there is still much debate and many unanswered questions. In this paper we will focus on one of these questions: if dark energy is due to a dynamical scalar field that can also have a non-gravitational interaction with dark matter, can this interaction affect the sound speed of dark energy?

When discussing exotic models of dark energy, the focus is often on the equation of state $w$ and how it affects observations such as the Cosmic Microwave Background (CMB) and the matter power spectrum. However, another parameter of interest is the dark energy speed of sound ($c_{\rm s}$) and its observational signatures and constraints using various cosmological probes (see, for example, \cite{Bean:2003fb, Weller:2003hw, Hannestad:2005ak, Xia:2007km, TorresRodriguez:2007mk, TorresRodriguez:2008et, Abramo:2009ne, Creminelli:2009mu, Anselmi:2011ef, Basse:2012wd, Appleby:2013upa, Batista:2013oca, Mehrabi:2015hva,Nesseris:2015fqa,Hojjati:2015qwa,Heneka:2017ffk, Batista:2017lwf}). 

Although the dark energy sound speed $c_{\rm s}$  remains practically unconstrained by observation, future cosmological experiments such as Euclid \cite{Amendola:2012ys} or SKA \cite{TorresRodriguez:2007mk, Maartens:2015mra} could constrain it. Therefore, it is important to fully explore the possibility of a varying $c_{\rm s}$ using well formulated, self consistent models and understand what effects this would have on various cosmological observables. This would affect the matter power spectrum through the growth of structure \cite{Anselmi:2011ef, Mehrabi:2015hva, Nesseris:2015fqa}, as well as the halo abundances and cluster counts \cite{Batista:2017lwf}. This is due to the fact that if $c_{\rm s}<1$ then dark energy clusters and affects observations in a non-trivial way.

Interacting dark energy models relax the assumption of $\Lambda$CDM and uncoupled dynamical dark energy models, which consider dark energy and dark matter to be only gravitationally coupled. In this paper we show that an interacting model of dark energy in the form of quintessence with evolving $w$ and $c_{\rm s}^2$ can be constructed from a Lagrangian. For this purpose we follow the Lagrangian formalism for coupled fluids developed in \cite{Pourtsidou:2013nha}, where three new general families of interacting quintessence and k-essence models were constructed. We focus in particular on their `Type 3' models, which are pure momentum-transfer models up to linear order in perturbation theory. We will examine the impact of such interactions on the CMB temperature and matter power spectra.

The key advantage to using the Lagrangian formalism, as opposed to an ad-hoc approach at the level of the field and fluid equations, is self consistency. This approach leads to dynamically evolving  $w$ and $c_{\rm s}^2$ that are directly derived from the Lagrangian. This approach also avoids unforeseen instabilities such as those discussed in \cite{Valiviita:2008iv}, since usual pathologies like ghost and strong coupling problems can be immediately identified from the Lagrangian. 

The plan of the paper is as follows: In Section \ref{sec:c_sDE} we summarise the general families of interacting dark energy models (i.e. Type 1, Type 2 and Type 3)  constructed in  \cite{Pourtsidou:2013nha}, and explore the properties of the dark energy sound speed for each one, demonstrating that Type 3 models are characterised by a varying dark energy sound speed. In Section \ref{sec:T3} we choose a specific Type 3 sub-case and derive the background and linear perturbations equations. We then evaluate the CMB and linear matter power spectrum for different values of the coupling using our modified version of the Einstein-Boltzmann solver \CLASS{}~\cite{Blas:2011rf}, and compare with uncoupled quintessence. We show and discuss the behaviour and effects of the coupling and the varying dark energy sound speed. {\cML Finally, we use a Markov chain Monte Carlo (MCMC) analysis to compare the chosen sub-case with $\Lambda$CDM using the Planck 2015 CMB data \cite{Ade:2015xua, Ade:2013tyw}}.  We conclude in Section \ref{sec:concl}. 

\section{\label{sec:c_sDE} The sound speed of dark energy coupled to dark matter}

We choose a Minkowski metric signature $(-+++)$ and begin by setting the speed of light, $c=1$. We write the Einstein field equations as
\begin{equation}
G_{\mu \nu}=8\pi G \bigg(T_{\mu \nu}^{({\rm SM})}+T_{\mu \nu}^{({\rm DM})}+T_{\mu \nu}^{({\rm DE})}\bigg) \, ,
\end{equation}
where $G_{\mu \nu}$ is the Einstein tensor, $G$ is Newton's constant, $T_{\mu \nu} $ is the energy momentum tensor, SM refers to the standard model particles and DM, DE to dark matter and dark energy, respectively.
The Bianchi identities imply that 
\begin{equation}
\nabla_\mu \bigg(T^{\mu ({\rm SM})} _{\;\; \nu}+T^{\mu ({\rm DM})} _{\;\; \nu}+T^{\mu ({\rm DE})} _{\;\; \nu}\bigg)=0 \, , 
\end{equation}
which describes the total energy-momentum conservation.  
For an uncoupled model, $\nabla_\mu T^{\mu \, (i)} _{\;\; \nu}=0$ for each species $i = ({\rm SM, DM, DE})$, and we assume that this is the case for standard model particles which is well supported by strong observational constraints on standard model interactions 
\cite{Carroll:1998zi}.  

For a model with a coupling between dark energy and dark matter, only their total energy-momentum is conserved.  That is, there exists a coupling current such that
\begin{equation}
\nabla_\mu T^{\mu ({\rm DM})} _{\;\; \nu}=-\nabla_\mu T^{\mu ({\rm DE})} _{\;\; \nu}=J_\nu \, .
\end{equation}
We will denote $\bar{J}_0 = Q$, $\delta J_0 = q$, and $\delta J_i = \nabla_i  S$, where bars signify background quantities \cite{Pourtsidou:2013nha}. Note that because of isotropy, $\bar{J}_i=0$.  

Using the relativistic fluid description, the energy-momentum tensor for a general perfect fluid is written as 
 \begin{equation}
 T_{\; \mu  \nu}= (\rho + P)U_{\mu} U_{\nu} + P g_{\mu \nu} \, ,
\end{equation} where $\rho$ is the energy density, $P$ the pressure, and $U^\mu$ the velocity of a general fluid.
The equation of state is then defined as $w \equiv \bar{P}/\bar{\rho}.$ 
The sound speed is defined by
\be
c^2_{\rm s} \equiv \frac{\delta P}{\delta \rho} \, ,
\ee where $\delta P$ is the pressure perturbation and $\delta \rho$ is the energy density perturbation in the fluid rest frame. 
Since it is defined in the rest frame of the fluid, this is a gauge invariant quantity. 

A Lagrangian formalism for models of dark energy in the form of a scalar field coupled to dark matter using the fluid description was developed in \cite{Pourtsidou:2013nha}, and we will utilise it to investigate the properties of the sound speed of dark energy in such models.
We begin by considering the general functional form of the Lagrangian for dark energy and dark matter \cite{Pourtsidou:2013nha}
\begin{equation}
L=L(n,Y,Z,\phi) \, ,
\end{equation}
where $\phi$ is the dark energy scalar field, $Y = \frac{1}{2}(\nabla_\mu \phi)^2$ the usual kinetic term \footnote{ In the literature this is commonly referred to as $X$, however we will use $Y$ to follow the notation in \cite{Pourtsidou:2013nha}.}, $Z=u^{\mu}\nabla_\mu \phi$ is a coupling of the dark matter fluid velocity $u^\mu$ to the gradient of the scalar field, and $n$ the dark matter fluid number density.
This Lagrangian can be used to discuss general classes of quintessence and k-essence dark energy models, including a non-gravitational coupling between dark energy and dark matter.  
By splitting the Lagrangian in different ways, the authors of \cite{Pourtsidou:2013nha} constructed three distinct families of coupled models. Here we will briefly review these types of models and concentrate on the implications of the interaction for the speed of sound of dark energy in the form of quintessence. 

\subsection{Uncoupled models}

We first consider models with no interactions, where the Lagrangian can be split into independent terms representing the dark energy and dark matter, 
\be
L =  F(Y,\phi) + f(n) \, . 
\ee
This class includes popular alternatives to the cosmological constant model, namely the k-essence and quintessence models. 

Quintessence models have a minimally coupled dynamical dark energy field.  These models have $F(Y,\phi) = Y + V(\phi)$, where $V(\phi)$ is the quintessence potential.  
Quintessence models have a variable $w$ and $c_{\rm s}^2=1$ so that dark energy perturbations are negligible.  One of the main goals of forthcoming large scale structure surveys like Euclid \cite{Amendola:2012ys} is determining the equation of state parameter $w$, since $w\neq-1$ would signal a deviation from the concordance model ($\Lambda$CDM). 

In the fluid description, $\bar{\rho}_\phi = \Fb-2\bar{Y} \Fb_Y $ and $\bar{P}_\phi=-\Fb$, where the subscript $Y$ denotes partial differentiation (i.e. $\Fb_Y \equiv \partial \Fb / \partial Y$). 
We also find \cite{Erickson:2001bq, Kunz:2012aw}
\be
c^2_{\rm s} = \frac{\Fb_Y}{\Fb_Y+2\Fb_{YY}\bar{Y}} \, .
\ee 
For a standard quintessence model, i.e. for $F(Y,\phi) = Y + V(\phi)$, we have $\Fb_Y = 1 $ and $\Fb_{YY} = 0$, so we see that the speed of sound is always equal to unity ($c^2_{\rm s} = 1$). 

However, for a class of models that contain non-canonical kinetic terms, termed k-essence models, we have $F(Y,\phi) = K(\phi)p(Y)$.  This results in a non-unitary speed of sound $c^2_{\rm s} \neq 1$ \cite{dePutter:2007ny}.

\subsection{\label{sec:type1} Interacting models (Types 1 and 2) }

Considering interacting models, \cite{Pourtsidou:2013nha} discussed different ways the general Lagrangian could be split, and constructed three general classes of coupled theories (labelled Type 1, 2 and 3).  
For Type 1 the Lagrangian takes the form 
\be
L=F(Y, \phi)+f(n,\phi) \, \label{eq:T1L},   
\ee 
with the $\phi$ dependence in $f$ leading to interactions between dark energy and dark matter.

By restricting $F = Y + V(\phi)$, we can describe general coupled quintessence models.  
Taking a particular form for $f(n,\phi) = g(n)e^{\alpha(\phi)}$, the coupling current was found to be  \cite{Pourtsidou:2013nha}
\be
J_\mu = -\rho \frac{\partial \alpha}{\partial \phi} \nabla_\mu \phi \, .
\ee 
 Considering a cold dark matter (CDM) fluid we can write $f=n e^{\alpha(\phi)}$ and by choosing a specific form $\alpha(\phi) = \alpha_0 \phi$ with $\alpha_0$ a constant, we recover one of the most commonly studied coupled quintessence models \cite{Amendola:1999er,Xia:2009zzb}. 

For the Type 1 models the background energy density and pressure are found to be \cite{Pourtsidou:2013nha}
\begin{align} \nonumber
\bar{\rho}_\phi &= \Zb^2 + \Fb, \\
\bar{P}_\phi &= - \Fb \, , 
\end{align}
 and the perturbed quantities are given by
\begin{align} \nonumber
 \delta \rho_{\phi} &= \Zb \delta Z +\frac{\partial F}{\partial \phi} \varphi  \, , \nonumber \\
\delta P_{\phi} &= \Zb \delta Z - \frac{\partial F}{\partial \phi} \varphi \, , \nonumber \\
\theta_{\phi} &= \frac{\varphi}{\dot{\bar{\phi}}} \, .
\end{align}
Here $\varphi \equiv \delta \phi$ is the field perturbation, and $\theta$ is the scalar mode of the momentum, such that $U_i = a\nabla_i \theta$ for a general fluid. Note that $\bar{Z} = -\dot{\bar{\phi}}/a$.
We also have \cite{Valiviita:2008iv}
\be
\delta P_\phi = c^2_{\rm s} \delta \rho_\phi + 3{\cal H}(c^2_{\rm s} - c^2_{\rm a})(1+w_\phi)\bar{\rho}_\phi \theta_\phi - (c^2_{\rm s} - w_\phi)Q\theta_\phi \, .
\ee
Here, $c_{\rm a}$ is the adiabatic sound speed with $c_{\rm a}^2= \dot{\bar{P}}_\phi/\dot{\bar{\rho}}_\phi$ \cite{Christopherson:2008ry}.
For a general Type1 model, we find that the dark energy sound speed is given by \cite{Skordis:2015yra}
\begin{equation}
\label{eq:cs2Type1}
c_{\rm s}^2=\frac{\Fb_Y}{\Fb_Y + 2\bar{Y} \Fb_{YY}} \, .
\end{equation} 
This is the same expression as in the case of uncoupled quintessence, which means that for Type 1 coupled quintessence we get the standard result $c^2_{\rm s}=1$. In the case of coupled k-essence we have $F = F(Y,\phi)$ and the speed of sound can be different than unity, as in the uncoupled k-essence case. 

Type 2 models interact instead through a coupling of the dark matter fluid velocity to the gradient of the scalar field; the Lagrangian is split as \cite{Pourtsidou:2013nha}
\be
L=F(Y, \phi)+f(n,Z) \, . \label{eq:T2L}
\ee
For CDM we can write $f(n,Z)=n h(Z)$, and for this case the coupling current is found to be \cite{Pourtsidou:2013nha}
\be
J_{\mu}=\nabla_\nu (\rho_c \beta u^\nu)\nabla_\mu \phi
\ee
where $\beta=h_Z/(h-Zh_Z)$ \cite{Skordis:2015yra}. The speed of sound for Type 2 models is also given by Equation~(\ref{eq:cs2Type1}) \cite{Skordis:2015yra}. 

\subsection{\label{sec:type3} Interacting models (Type 3) }
Type 3 models are classified by the Lagrangian \cite{Pourtsidou:2013nha}
\begin{equation}
L = F(Y,Z,\phi)+f(n),  
\end{equation} 
where again  $Z \equiv u^\mu \nabla_\mu \phi$ couples the dark matter fluid velocity to the gradient of the scalar field. {\cML 
Here the coupling current is
\be
J_{\mu} = q^\beta_{\; \mu} \Big(\nabla_\nu (F_Z u^\nu)\nabla_\beta \phi 
+ F_Z \nabla_\beta Z + Z F_Z u^\nu \nabla_\nu u_\beta \Big) \, ,
\ee 
with 
$q^\nu_{\; \mu} =  u^\nu u_\mu +\delta^\nu_{\; \mu}$. 
From the above formula we calculate $J_0 = 0$ up to second order. This means $Q=q=0$ for all Type 3 models, but $\delta J_i \equiv \nabla_i S \neq 0$, so Type 3 is a theory of pure momentum exchange up to linear order \cite{Pourtsidou:2013nha}. }

The background energy density and pressure are
\begin{align} \nonumber
\bar{\rho}_{\phi} &= \bar{Z}^2 \bar{F}_Y-Z \bar{F}_Z +\bar{F} ,\\
\bar{P}_{\phi} &= -\bar{F}, \label{eq:rhoP}
\end{align}
and the perturbed quantities are 
\begin{align} \nonumber
\delta \rho_{\phi} &= \bar{Z}[\bar{F}_Y-\bar{Z}^2\bar{F}_{YY} +2\bar{Z} \bar{F}_{Y Z} -\bar{F}_{ZZ}] \delta Z\\&+[\bar{Z}^2 \bar{F}_{Y \phi} - \bar{Z}\frac{\partial \Fb_Z}{\partial \phi} +\frac{\partial \Fb}{\partial \phi}]\varphi \, , \nonumber \\ 
\delta P_{\phi} &= (\Zb \Fb_Y - \Fb_Z)\delta Z -\frac{\partial \Fb}{\partial \phi} \varphi \, , \nonumber \\ 
\theta_{\phi} &= \frac{a^{-1}\bar{F}_Y\varphi +\bar{F}_Z \theta_{c}}{\bar{F}_Z-\bar{Z}\bar{F}_Y} \, .
\end{align} 
In these models the sound speed is found to be \cite{Skordis:2015yra}
\begin{equation}
c_{\rm s}^2=\frac{\bar{Z}\bar{F}_Y-\bar{F}_Z}{\Zb (\Fb_Y-2\Zb \Fb_{ZY}-\Fb_{ZZ}-\Zb^2\Fb_{YY})} \, .
\label{eq:c2full}
\end{equation}
Assuming a quintessence form for $F$ we can write
\begin{equation}
F = Y+V(\phi)+h(Z) \, .
\end{equation}
We then find
\begin{equation}
c_{\rm s}^2=\frac{\Zb-h_Z}{\Zb(1-h_{ZZ})} \, . \label{eq:c2quin}
\end{equation}

Type 3 models with a quadratic coupling, i.e. $h(Z) = \beta_0 Z^2$ with $\beta_0$ a dimensionless coupling constant, have been already proven to be phenomenologically interesting, since it has been shown that they can reconcile the $\sigma_8$ tension between high and low redshift cosmological probes \cite{Pourtsidou:2016ico}.
 For this case the dark energy sound speed is equal to unity as
\be
c^2_{\rm s} = \frac{1-2\beta_0}{1-2\beta_0} = 1 \, .
\ee
Here we will generalise the coupling function to $h(Z) = \beta_{n-2} Z^n$ with $n$ an integer $n \geq 2$; this can still be thought of as similar to the quadratic case, but allowing for a $Z$-dependent (hence time dependent) dimensionless coupling parameter $\beta(Z) = \beta_{n-2} Z^{n-2}$. 
We find
\be
c^2_{\rm s} = \frac{1-\beta_{n-2} n \Zb^{n-2}}{1-\beta_{n-2}n (n-1) \Zb^{n-2}} \, , \label{eq:c2T3}
\ee with $\Zb = -\dot{\bar{\phi}}/a$. 

From this relation it is evident that the speed of sound can deviate from unity for $n \neq 2$. In the limit $|\beta_{n-2} \bar{Z}^{n-2}| \gg 1$ we find
\be
c^2_{\rm s} \rightarrow \frac{1}{n-1} \, .
\label{eq:c2slim}
\ee
This means, that in the case where the coupling is large, $c_{\rm s}^2$ reaches a constant value. 

In the following section we will implement a Type 3 model with variable speed of sound in the Einstein-Boltzmann solver \CLASS{}~\cite{Blas:2011rf}, which will allow us to quantify the above properties.

\section{Interacting model with variable speed of sound} \label{sec:T3}
In the previous section we showed that it is possible to construct a Lagrangian describing a quintessence field coupled to dark matter that results in an effective dark energy sound speed deviating from unity. In this section we will study the phenomenology of such models by exploring a specific case
\begin{equation}
F = Y+V(\phi)+h(Z) \, ,
\end{equation} with $h(Z) = \beta_1 Z^3$, where the effective dimensionless coupling parameter is $\beta_1 Z$.

Following \cite{Pourtsidou:2013nha} we 
will work in the synchronous gauge and 
assume a Universe described by a flat Friedmann-Lema\^{i}tre-Robertson-Walker (FLRW) metric 
\be
ds^2 = a^2(\tau)(-d\tau^2 + dx_i dx^i)
\ee in the background, and 
\be
ds^2 = -a^2 d\tau^2 + a^2 \left[\left(1+\frac{1}{3}h\right)\gamma_{ij}+ D_{ij}\nu\right]dx^idx^j \, 
\ee for linear perturbations, where $\gamma_{ij}$ is the metric for a 3 dimensional, spatial hyper-surface,   $\vec{\nabla}_k$ is the covariant derivative associated to $\gamma_{ij}$ such that, $\vec{\nabla}_k \gamma_{ij}=0$  and $D_{ij}$ is the traceless derivative operator $D_{ij}=\vec{\nabla}_i\vec{\nabla}_j-\frac{1}{3}\vec{\nabla}^2 \gamma_{ij}$. 

The scalar field action in the dark matter frame is 
\be
S_\phi=\int dt d^3 x a^3 \bigg[\frac{1}{2}(1+2 \beta_1 \frac{\dot{\phi}}{a})\dot{\phi}^2-\frac{1}{2}|\vec{\nabla} \phi|^2 -V(\phi)\bigg] \, .
\label{eq:action}
\ee
In the models we consider, we will have $\dot{\phi} > 0$ and so we limit ourselves to positive values of $\beta_1$; this ensures that we do not have to worry about ghosts or strong coupling pathologies in the model.  In addition, Equation~(\ref{eq:c2quin}) suggests that there is a singularity in $c_{\rm s}^2$ if $h_{ZZ}=-6\beta_1\dot{\bar{\phi}}/a\rightarrow 1$, but for positive values of $\beta_1$ we have $h_{ZZ} <0$ so this potential instability does not manifest itself.

For the Type 3 model under consideration the background energy density and pressure for the field follow from Equation~(\ref{eq:rhoP}):
\begin{align} \nonumber
\bar{\rho}_{\phi} &= \left(\frac{1}{2}+2\beta_1\frac{\dot{\bar{\phi}}}{a}\right)\frac{\dot{\bar{\phi}}^2}{a^2}+V(\phi) \, ,\\
\bar{P}_{\phi} &=\left(\frac{1}{2} +\beta_1\frac{\dot{\bar{\phi}}}{a}\right)\frac{\dot{\bar{\phi}}^2}{a^2}-V(\phi)\, .
\end{align}
The perturbed quantities are found to be
\begin{align} \nonumber
\delta \rho_{\phi} &= \frac{\dot{\bar{\phi}}}{a^2}\left(1 +6\beta_1 \frac{\dot{\bar{\phi}}}{a}\right)\dot{\varphi}+\frac{\partial V}{\partial \phi} \varphi  \, ,\nonumber \\ 
\delta P_{\phi} &= \frac{\dot{\bar{\phi}}}{a^2}\left(1  +3\beta_1 \frac{\dot{\bar{\phi}}}{a}\right)\dot{\varphi}-\frac{\partial V}{\partial \phi} \varphi \,  ,\nonumber \\ 
\theta_{\phi} &= \frac{a\varphi +3\beta_1\dot{\bar{\phi^2}}\theta_{c}}{3\beta_1\dot{\bar{\phi^2}}+a \dot{\bar{\phi}}} \, .
\end{align} 
For this model the sound speed follows from (\ref{eq:c2T3}):
\begin{equation}
\label{eq:cs2T3}
c_{\rm s}^2=\frac{1+3\beta_1(\dot{\bar{\phi}}/a)}{1+6\beta_1(\dot{\bar{\phi}}/a)} \, .
\end{equation} 
We also derive the background
\be
\left(1+6\beta_1\frac{\dot{\bar{\phi}}}{a}\right)(\ddot{\bar{\phi}}-\mathcal{H}\dot{\bar{\phi}})
+3\mathcal{H}\dot{\bar{\phi}}(1-3\beta_1\Zb)+a^2 V_\phi = 0
\ee
and perturbed Klein-Gordon equations 
\begin{align} \nonumber
\left(1+6\beta_1\frac{\dot{\bar{\phi}}}{a}\right)(\ddot{\varphi}+2\mathcal{H}\dot{\varphi})-6\beta_1\dot{Z}\dot{\varphi}+(k^2+a^2 V_{\phi \phi})\varphi \\
+(\dot{\bar{\phi}}+3\beta_1\frac{\dot{\bar{\phi}}^2}{a})\frac{\dot{h}}{2}+3 k^2 \beta_1\frac{\dot{\bar{\phi}}^2}{a} \theta_{c} = 0.
\end{align}

\subsection{Initial conditions}

In order to study the observational signatures of this model, we implement it in the Einstein-Boltzmann solver \CLASS{} \cite{Blas:2011rf}.
We assume a single exponential potential
\begin{equation}
V(\phi)=V_0 e^{-\lambda \phi} \, ,
\end{equation}
with $\lambda = 1.22 \, [m_{\rm Pl}]^{-1}$ where $m_{\rm Pl}$ is the reduced Planck mass; the parameter $\lambda$ can have a range of values and we choose this value as it leads to uncoupled and coupled quintessence models consistent with observations \cite{Xia:2007km, Pourtsidou:2016ico}; $V_0$ is tuned by the code to match a fixed $\Omega_\phi$ today. The initial conditions for the quintessence field are chosen to be $\phi_{\rm ini} = 10^{-4}  \, [m_{\rm Pl}]$ and $\dot{\phi}_{\rm ini}$ = $0$, initially, as in \cite{Pourtsidou:2016ico}. The dynamics then quickly approaches the tracking solution.

It is useful to discuss how the dark energy perturbations are initialised. We focus on the uncoupled case for simplicity, though this should not make much difference as at early times the coupling is small (see Fig. \ref{fig:betaZ}). We work in the synchronous gauge and follow the description in \cite{Ballesteros:2010ks} (note our conventions are different due to the difference in definitions of $\theta_i$). Starting from the continuity and Euler equations for a fluid without anisotropic stress, we find 
\begin{align}
\dot{\delta}_i=&-(1+w)(\nabla^2 \theta_i+\frac{\dot{h}}{2})-3(c_{\rm s}^2-w)\mathcal{H}\delta_i \nonumber \\
&-9(1+w)(c_{\rm s}^2-c_{\rm a}^2)\mathcal{H}^2\theta_i, \\
\dot{\theta}_i=&-(1-3c_{\rm s}^2)\mathcal{H}\theta_i+\frac{c_{\rm s}^2 }{1+w}\delta_i \, ,
\end{align} 
where the subscript $i$ refers to the different species.

We assume for simplicity a constant equation of state and sound speed.  
As we initialise our numerical studies from early times, we work in the  radiation dominated era, providing simple relations for ${\cal H}$, $a$ and $\tau$. We will also use that $h\propto (k\tau)$ outside the horizon.
We can then derive a relation between the metric perturbations $h$ and $\delta_{i}$, outside of the horizon to leading powers in $k\tau$: 
\begin{align}  
\delta_i&=-(1+w)  \frac{(4-3c_{\rm s}^2)}{(4-6w+3c_{\rm s}^2)} \frac{h}{4}\label{eq:initaldeltasD} \,  ,\\
\theta_i&=-  \frac{c_{\rm s}^2 \tau}{(4-6w+3c_{\rm s}^2)}\frac{h}{4}  \label{eq:initaldeltasT} \, .
\end{align}
Reproducing this in the matter dominated era gives:
\begin{align}  
\delta_i&=-(1+w)  \frac{(5-6c_{\rm s}^2)}{(5-15w+9c_{\rm s}^2)} \frac{h}{2}\label{eq:initaldeltasDMatter} \,  ,\\
\theta_i&=-(1+w)  \frac{\tau c_{\rm s}^2}{(5-15w+9c_{\rm s}^2)} \frac{h}{2}\label{eq:initaldeltasTMatter} \, .
\end{align}

\begin{figure}[H]
  \centering
  \includegraphics[width=0.7\textwidth]{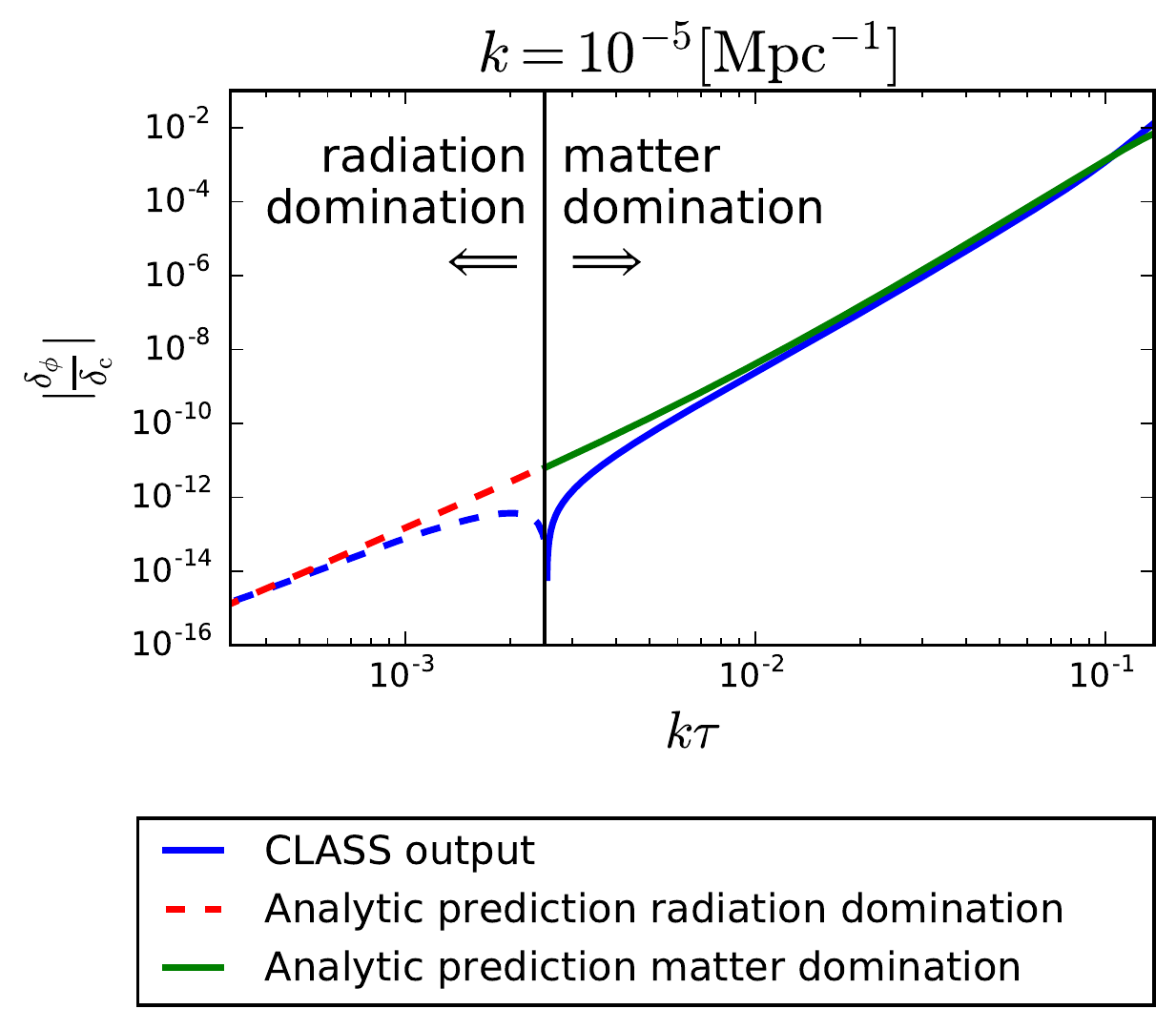}
  \caption{The plot shows how the analytic predictions (red and green lines) compares with the \CLASS{} output (solid line shows where  $\frac{\delta_\phi}{\delta_{\rm c}}<0$ and the dashed lines shows $\frac{\delta_\phi}{\delta_{\rm c}}>0$) for $k=10^{-5} \, {\rm Mpc}^{-1}$.  This value of $k$ is chosen such that $k\tau<1$ for the age of the universe and thus we can trust the first order result given in Equation~(\ref{eq:initaldeltasD}), but the behaviour remains qualitatively the same for any value of $k$ given this condition is met. The analytic solutions are derived using the values from \CLASS{} for $c_{\rm s}^2$ and $w$.} 
\label{fig:ICs}
\end{figure}

Although these initial conditions assume constant $w$ and $c_{\rm s}^2$ (as we are focused on the uncouple model $c_{\rm s}^2=1$), the analytic predictions are in reasonable agreement with the output from \CLASS{}.
This is demonstrated in Fig. \ref{fig:ICs}.  Even when the initial conditions are away from this solution the expected behaviour is quickly found and this does not have a significant effect on the late time result for the dark energy perturbations. Note the change in sign in $\delta_{\phi}$ when using \CLASS{}; this occurs as the model evolves from the radiation dominated era to matter domination, which can also be seen when comparing the analytic solutions Equation~(\ref{eq:initaldeltasD}) and (\ref{eq:initaldeltasDMatter}). 

\subsection{Sound speed evolution}

\begin{figure}[H]
  \centering
  \includegraphics[width=0.7\textwidth]{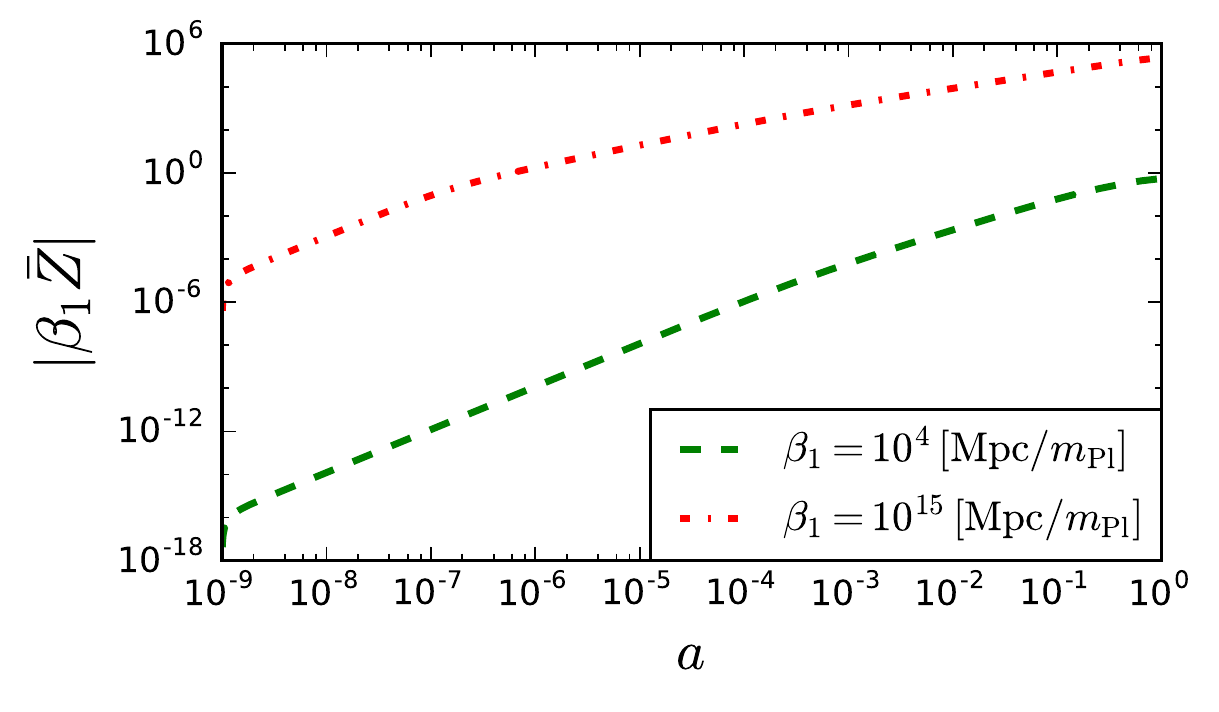}
  \caption{The evolution of the dimensionless, time-dependent coupling $\beta_1 \Zb$ for two values of $\beta_1$.}
  \label{fig:betaZ}
\end{figure}

The sound speed behaviour reflects how the dimensionless quantity $\beta_1 \Zb$ evolves over time, shown in Fig.~\ref{fig:betaZ}. 

\begin{figure}[H]
  \centering
  \includegraphics[width=0.7\textwidth]{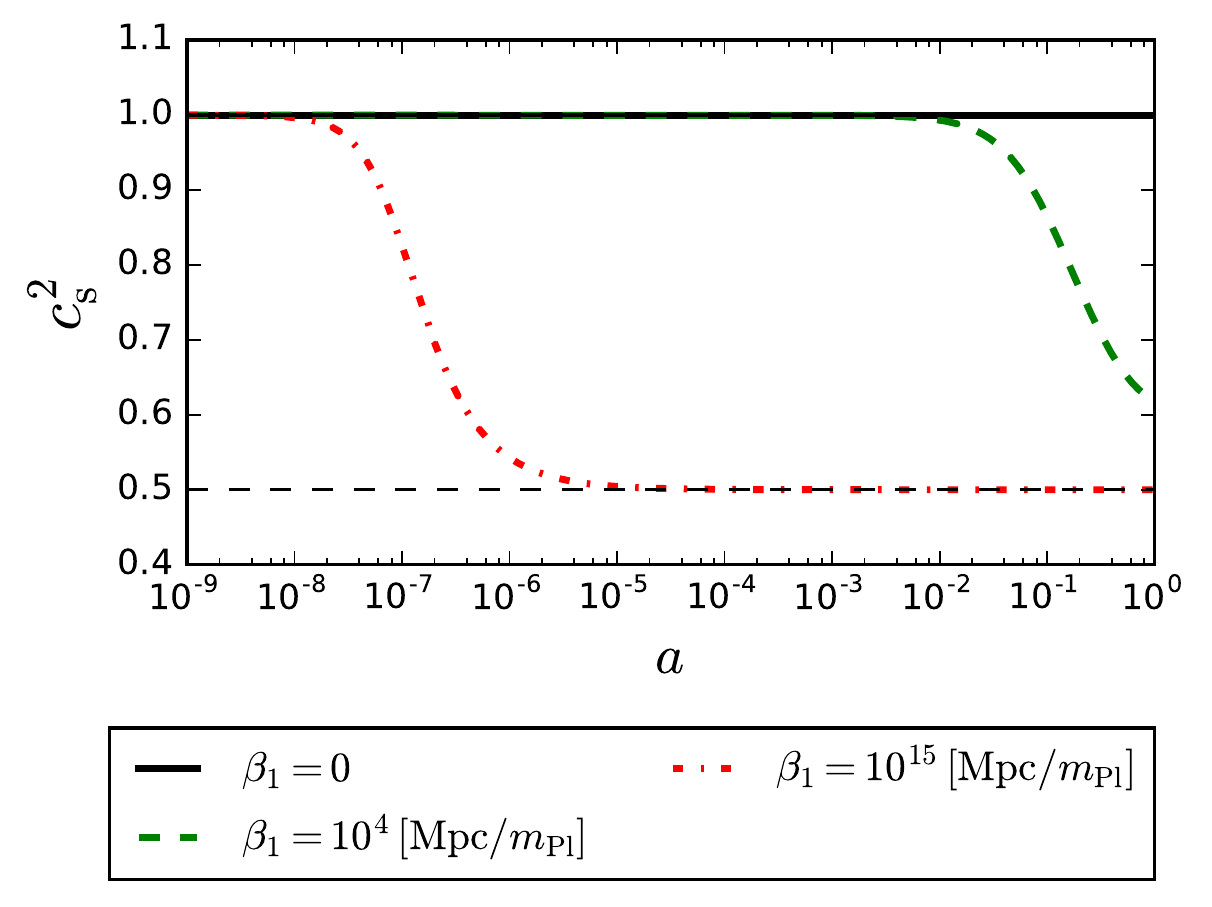}
  \caption{The evolution of $c_{\rm s}^2$ for the uncoupled case $\beta_1 = 0$ (solid black line) and two coupled cases (dot-dashed and dashed coloured lines). The dashed black line shows the limit $c^2_{\rm s} = 1/2$. }
  \label{fig:cs2}
\end{figure}

The coupling behaviour is directly reflected in the evolution of the sound speed, shown in Figure~\ref{fig:cs2}. 
From Equation~(\ref{eq:cs2T3}), we see that the sound speed can vary from unity for $|\beta_1 \Zb| \ll 1 $ to  $c_{\rm s}^2 \rightarrow\frac{1}{2}$ for $|\beta_1 \Zb| \gg 1$.   This is reflected in the figures: $|\beta_1 \Zb| \ll 1$ at early times and the sound speed is unity, while at later times its increase translates to $c^2_{\rm s} < 1$. 
For larger couplings, the deviation from $c_{\rm s}^2=1$ occurs at earlier times. 

\begin{figure}[H]
  \centering
  \includegraphics[width=0.7\textwidth]{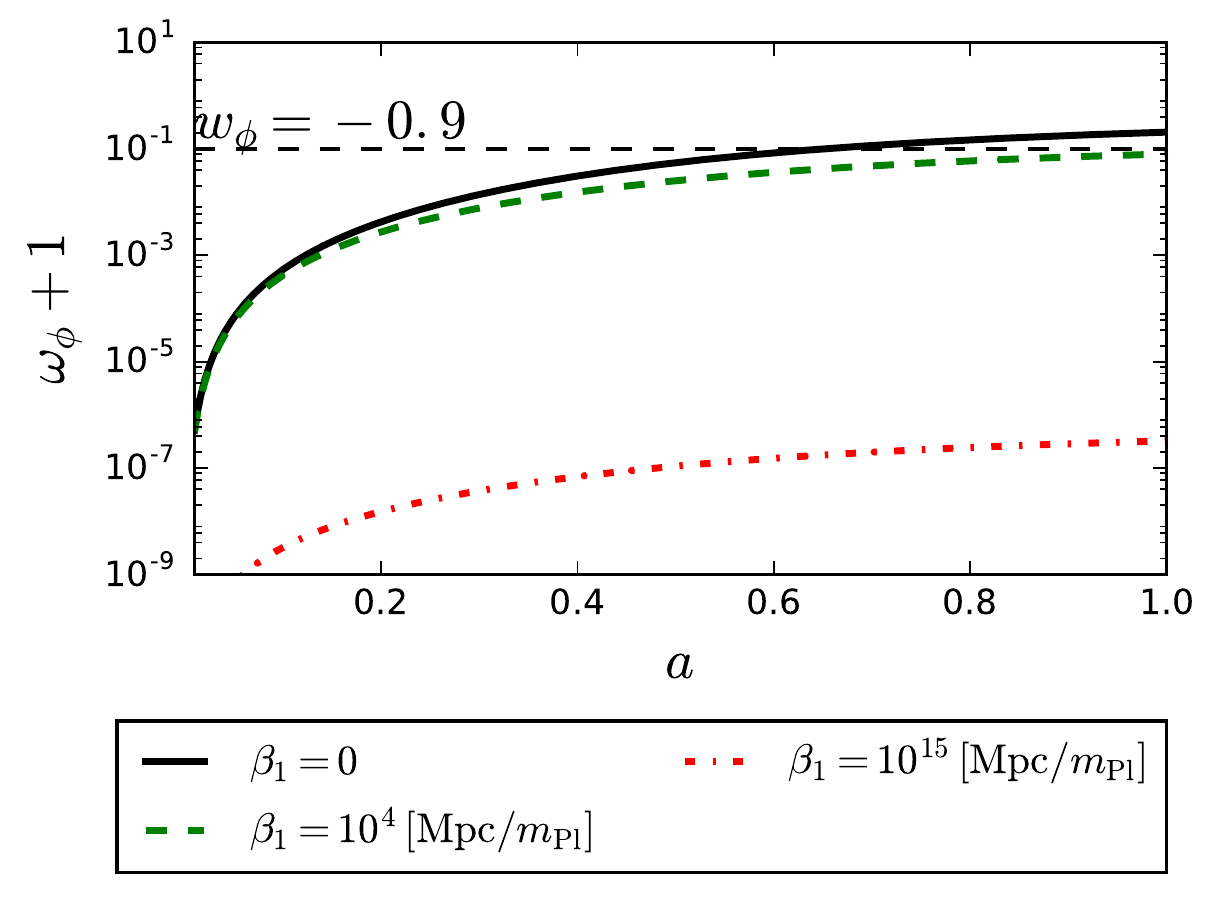}
  \caption{The evolution of the equation of state $w_\phi = \bar{P}_\phi / \bar{\rho}_\phi$ as a function of the coupling parameter $\beta_1$. A constant $w_\phi = -0.9$ is shown for comparison.}
  \label{fig:w}
\end{figure}

We have also included a plot of the equation of state $w_\phi = \bar{P}_\phi / \bar{\rho}_\phi$ in Figure~\ref{fig:w}. For the largest coupling it remains practically constant and very close to $-1$, while for the smaller coupling it evolves considerably and is more similar to the $w_\phi$ of uncoupled quintessence. This agrees with the findings in \cite{Pourtsidou:2016ico} and it is because the term $\propto \beta_1$ in the equation of state formula becomes completely subdominant to $V(\phi)$ and $w_\phi \rightarrow -1$.

In \cite{Bean:2003fb} it was shown that the effects of a non-unitary sound speed are more pronounced when $(w+1)$ is large. This suggests that its effects are suppressed for large positive $\beta_1$ and that any observable effects of the non-unitary sound speed would appear only at late times. 

We next examine the effects that this new type of interaction has on cosmological observables, focusing on the CMB temperature (TT) and matter power spectra. In order to highlight the effects of the coupling we fix the sound horizon angular scale at decoupling $\theta_{\rm s}$ and the physical energy densities of CDM and baryons, $\omega_{\rm c,b} = \Omega_{\rm c,b}h^2$ to the Planck 2015 best-fit model \cite{Ade:2015xua}. 

\subsection{CMB}

\begin{figure}[H]
  \centering
  \includegraphics[width=1\textwidth]{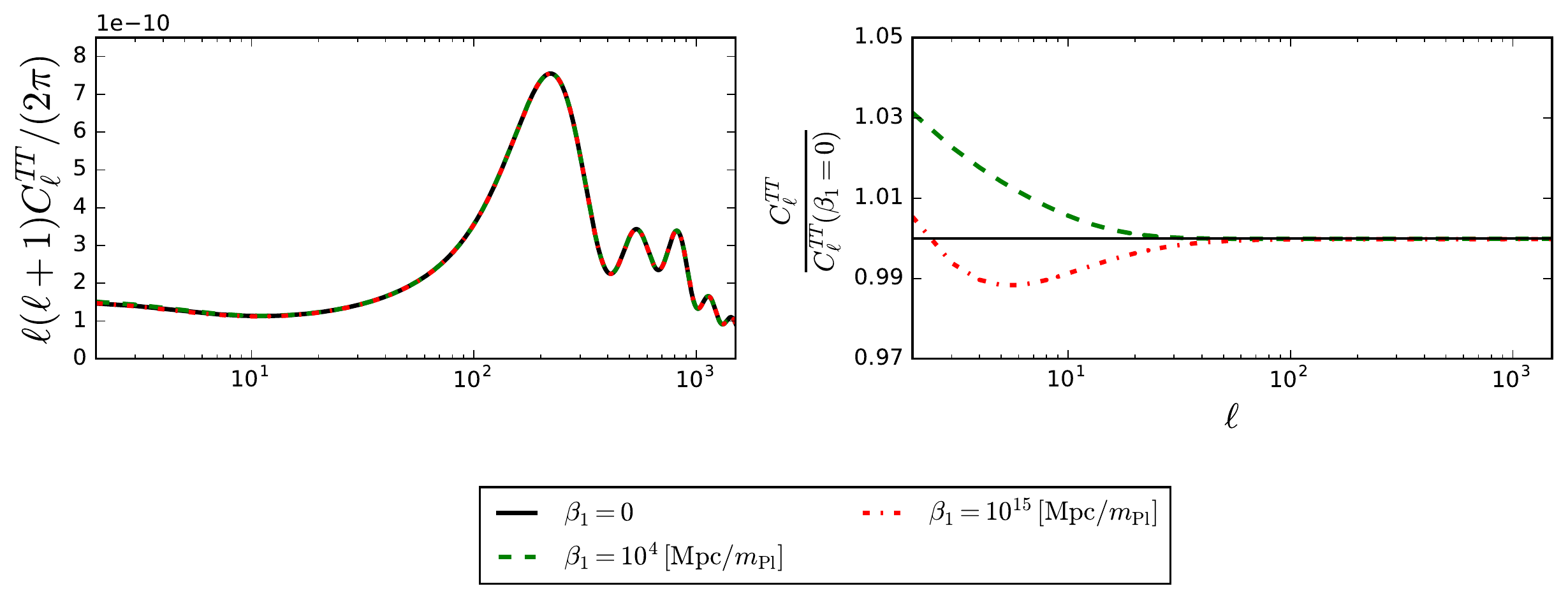}
  \caption{ Comparison of the CMB temperature (TT) power spectra for a range of values of $\beta_1$. The plots show the predictions from the coupled models and uncoupled quintessence (Left) as well as the ratio between the coupled models and uncoupled quintessence (Right).}
  \label{fig:CMB}
\end{figure}

In Figure~\ref{fig:CMB} (Left) we show the CMB temperature power spectra for the chosen coupled models and the predictions of the uncoupled quintessence model, as well as the ratio of the coupled models CMB spectra to the uncoupled one (Right). We can see that the greatest impact is on the largest scales through the Integrated Sachs-Wolfe (ISW) effect. These differences are small relative to the large cosmic variance and the theories remain consistent with current observational data. Similar effects were seen in the quadratic coupling case \cite{Pourtsidou:2016ico}. 

\subsection{Matter power spectrum}

\begin{figure}[H]
  \centering
  \includegraphics[width=1\textwidth]{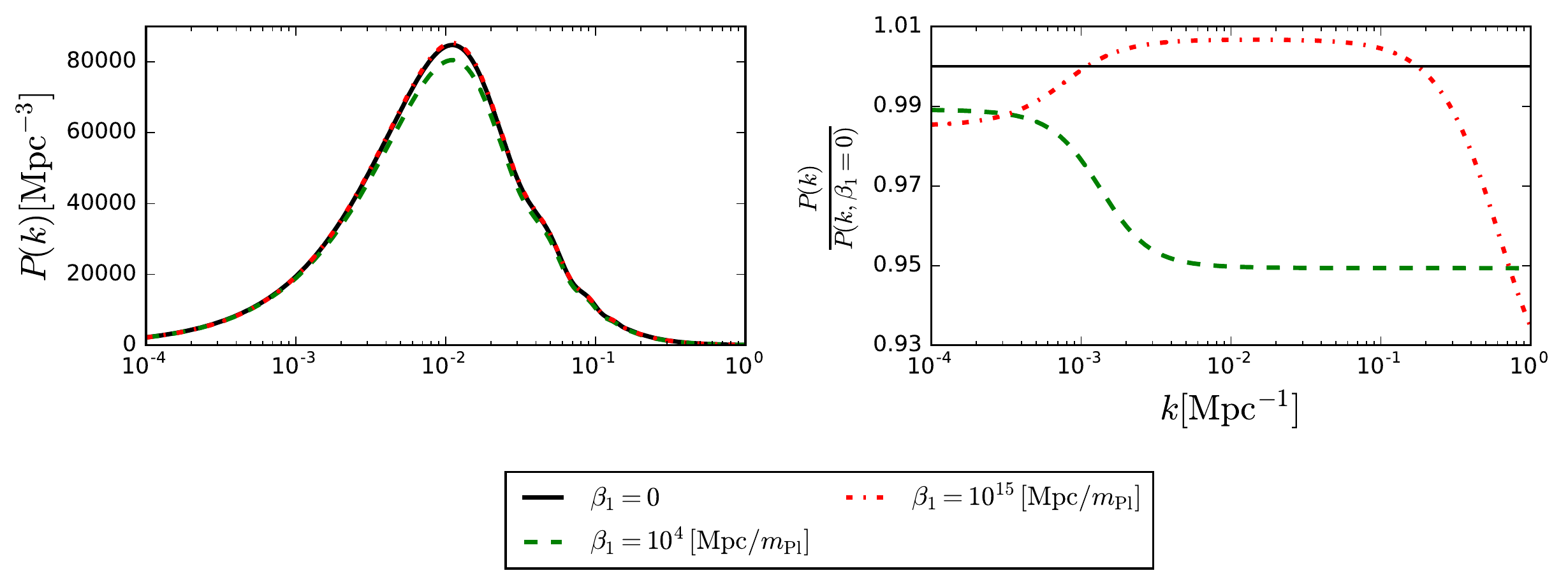}
  \caption{Comparison of the linear matter power spectrum $P(k)$ at $z=0$ for a range of values of the coupling parameter $\beta_1$. The plots show the predictions from the coupled and uncoupled quintessence models (Left) as well as the ratio between the coupled models and uncoupled quintessence (Right). Note that the linear description is expected to be valid only up to $k \sim 0.1 \, {\rm Mpc}^{-1}$.}
  \label{fig:mat}
\end{figure}

{\cML In Figure~\ref{fig:mat} (Left) we show the matter power spectra at $z=0$ for the coupled models and the predictions of the uncoupled quintessence model, as well as the ratio of the coupled models matter power spectra to the uncoupled one (Right). 
Note that we chose to show the matter power spectra in log-linear scale instead of the traditional log-log scale, because the differences are clearer using the former. }
We see a suppression of power on observable scales for $\beta_1 = 10^4 [{\rm Mpc}/m_{\rm Pl}]$ and an increase for the large value of the coupling parameter  $\beta_1 = 10^{15}[{\rm Mpc}/m_{\rm Pl}]$.

This feature has been also seen in other momentum transfer interacting dark energy models \cite{Pourtsidou:2016ico}; in the same paper it was shown that the models that suppress power can reconcile the $\sigma_8$ tension between CMB and large scale structure data.

\begin{figure}[H]
  \centering
  \includegraphics[width=0.7\textwidth]{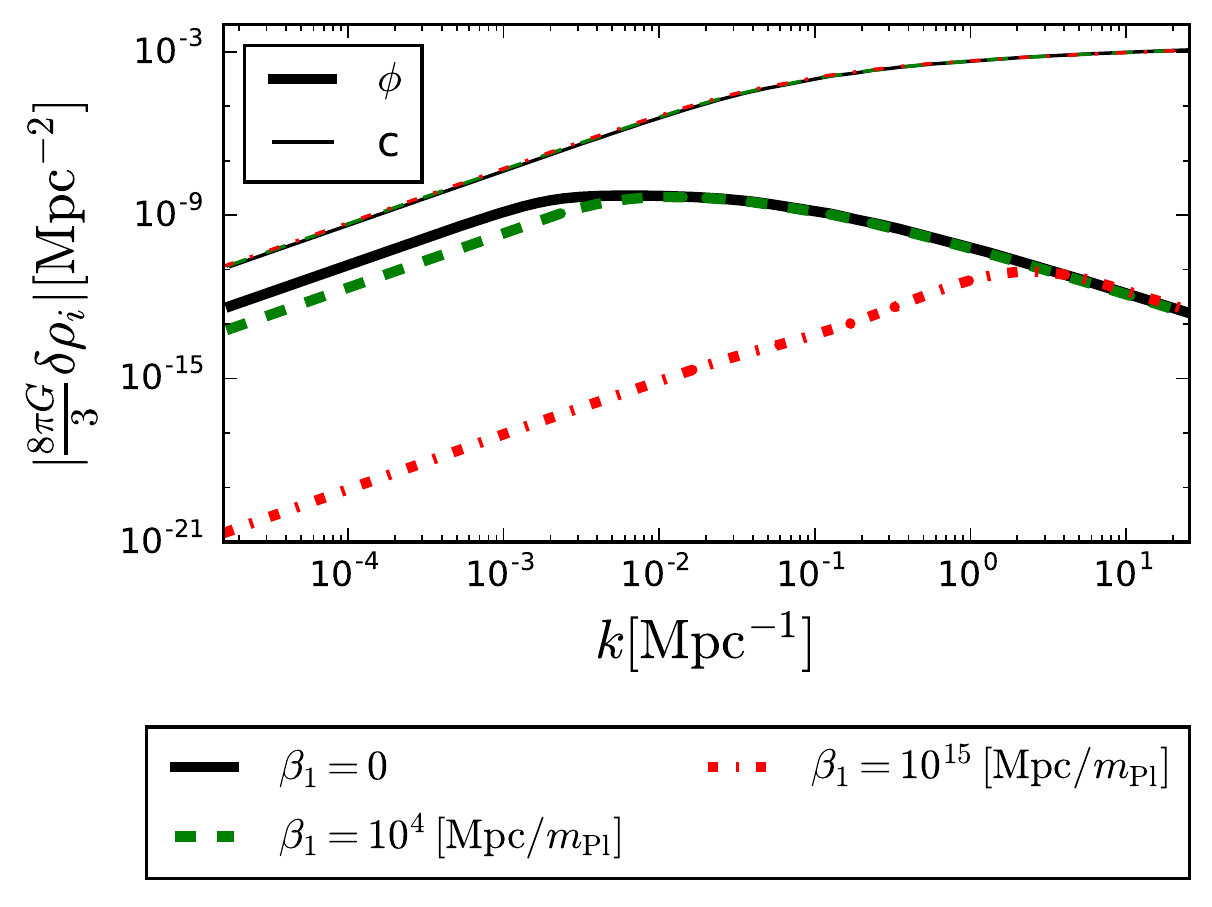}
  \caption{The evolution of $\delta \rho_{\phi}$ (thick lines) and  $\delta \rho_{c}$ (thin lines) as a function of $k$ for a variety of couplings. Note that the linear description is expected to be valid only up to $k \sim 0.1 \, {\rm Mpc}^{-1}$.}
  \label{fig:pert}
\end{figure}

It is also useful to investigate how the changing sound speed affects the dark energy power spectrum; we do this by comparing the size of the perturbations, $\delta \rho_{\phi}$ and $\delta \rho_{c}$, for different values of the coupling; this is shown in  Figure~\ref{fig:pert}. One can see that the variable sound speed should make very little difference to the \emph{total} (CDM+DE) power spectra for our models, since the dark energy perturbations remain much smaller than the CDM ones for all scales and couplings. In \cite{Sergijenko:2014pwa} it is shown how changing the $c_{\rm s}^2$ in an uncoupled quintessence model affects the behaviour of the dark energy perturbations. As $c_{\rm s}^2\rightarrow 0$ they find an increase in $\delta_\phi(k)$ for all values of $k$, however it is still the case that $|\delta_\phi(k)| \ll |\delta_{\rm c}(k)|$. Comparing Figure~\ref{fig:pert} to the results in \cite{Sergijenko:2014pwa} implies that in this coupled model $\delta_\phi$ is insensitive to the changes we observe in $c^2_{\rm s}$, which is expected as in our case $c^2_{\rm s}$ never becomes very small. However, the analytic prediction given in Equation~(\ref{eq:initaldeltasD}) suggests $\delta_\phi$ is very sensitive to $w_\phi$. In Figure~\ref{fig:w} one can see that $w_\phi$ is sensitive to the coupling. This is the dominant effect that leads to $\delta_\phi$ varying with $\beta_1$, rather than the change we see in the sound speed.

\subsection{MCMC analysis}
{\cML In this Section we perform a first MCMC analysis using the TT (including the low $\ell$) and the lensing reconstruction from Planck 2015 CMB data set \cite{Ade:2015xua, Ade:2013tyw}, and the  \MontePython{} code \cite{Audren:2012wb}, and compare with $\Lambda$CDM. A full analysis using a suite of CMB and low-$z$ data sets is left for an upcoming publication, where we will also study other Type 3 models.

We exclude the negative values of $\beta$ as they can lead to pathologies like ghosts (see Equation~(\ref{eq:action})) and choose the following priors for $\lambda$ and $\beta_1$:
\be 
\lambda \in [0; 2.1],  
\qquad 
\log_{10} \beta_1  \in [-4; 15] \, .
\ee
We choose flat priors for the rest of the cosmological parameters ($\omega_{\rm b}, \omega_{\rm cdm}, \theta_{\rm s}, A_{\rm s}, n_{\rm s}, \tau_{\rm reio}$), and the collection of nuisance parameters required by the Planck likelihoods.  

\begin{table}[H]
\centering
\renewcommand{\arraystretch}{1.2}

\begin{tabular}{ |p{2cm}|p{3.2cm}|p{3.1cm}|  }
 \hline
& $\Lambda$CDM &T3 [$h(Z)=\beta_1 Z^3$]\\
 \hhline{|=|=|=|}
 $100~\omega_{\rm b }$ & $2.23_{-0.02}^{+0.02}$ & $2.23_{-0.02}^{+0.02}$\\
$\omega_{\rm cdm }$ & $0.119_{-0.002}^{+0.002}$& $0.119_{-0.002}^{+0.002}$\\
$10^9 A_{\rm s}$&$2.16_{-0.06}^{+0.07}$&$2.16_{-0.07}^{+0.05}$\\
$n_{\rm s }$ & $0.967_{-0.006}^{+0.006}$&$0.967_{-0.006}^{+0.006}$\\
$\tau_{\rm reio }$ & $0.07_{-0.02}^{+0.02}$ & $0.07_{-0.02}^{+0.02}$\\
$\sigma_8$ &  $0.818_{-0.010}^{+0.010}$ & $0.795_{-0.014}^{+0.032}$ \\
 $H_0$   & $67.8_{-0.9}^{+0.9}$&$66.8_{-0.6}^{+2.2}$\\
 $\lambda$ &   -  & $0.9_{-1}^{+0.3}$   \\
 $\log_{10} \beta_1 $&- & $6.0_{-10}^{+9}$ \\
 $\chi^2$&$11271.38$&$11271.80$ \\
 \hline
\end{tabular}
\renewcommand{\arraystretch}{1}
 \caption{Cosmological parameters for $\Lambda$CDM and the T3 model with the coupling function $h(Z)=\beta_1 Z^3$, including the $\chi^2$ value.}

\end{table}

\begin{figure}[H]
  \centering
  \includegraphics[width=1\textwidth]{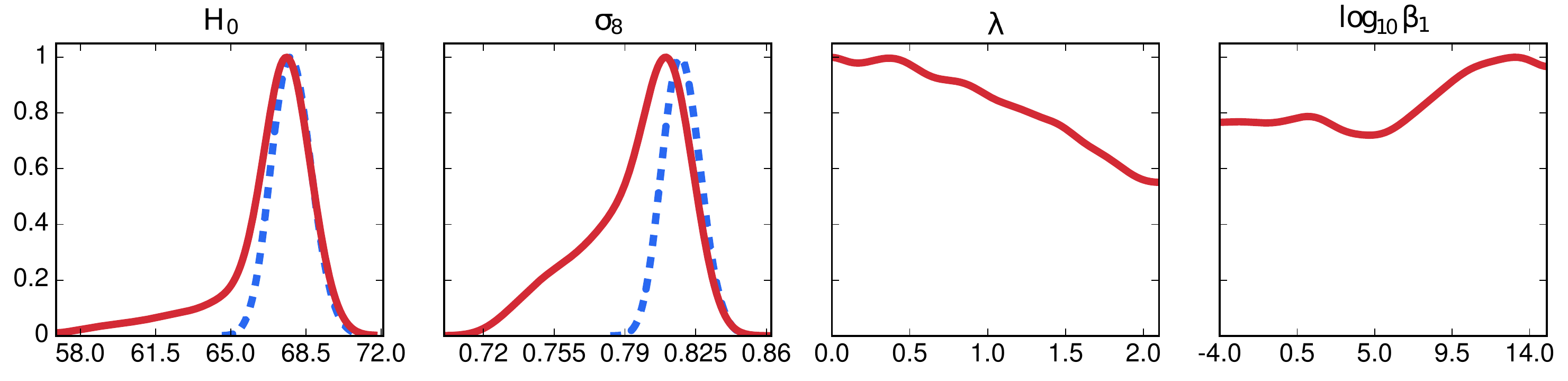}
  \caption{One-dimensional posterior distributions of the parameters $\{\sigma_8, H_0, \beta_1, \lambda\}$ for the coupled T3 model (solid red lines) and \Lcdm{} (blue dashed lines).}
   \label{fig:MCMC}
\end{figure}

This analysis shows that this model is compatible with the current constraints from Planck, though marginally disfavoured if compared to $\Lambda$CDM, additionally a full Bayesian analysis would disfavour the T3 model due to the extra parameters. This is qualitatively similar with the results in \cite{Pourtsidou:2013nha}, where it is shown that the advantage of T3 models is that they can resolve the current (tentative) $\sigma_8$ tensions between high and low redshift data. Examining this in detail for a variety of momentum transfer models will be the subject of an upcoming publication. We expect that adding low redshift data will help mitigate the $\sigma_8$ tension and constrain $\beta_1$, as seem in \cite{Pourtsidou:2013nha}.}

\section{Conclusions and Discussion} 
\label{sec:concl}

In this paper we have presented a new approach to exploring the sound speed of dark energy. Using the Lagrangian formalism, we demonstrated how one can obtain a dark energy quintessence field with varying sound speed via pure momentum exchange interactions with dark matter. We also examined the effect this kind of interaction has on cosmological observables, such as the CMB temperature and matter power spectra, {\cML and showed that the model is compatible with current cosmological constraints from the Planck mission}. 

For the most common coupled quintessence models (Types 1 and 2), we have shown that the speed of sound is always unity ($c_{\rm s}^2=1$). For Type 3 quintessence models that involve a coupling of the fluid velocity to the gradient of the scalar field, one can easily construct a model with an evolving sound speed. Type 3 models are special as the form of the coupling results in an effective ``non-canonical'' kinetic term, similar to k-essence, which then allows for a varying sound speed of dark energy. 

Using our modified version of \CLASS{} we have looked at the impact of such an interaction in the CMB temperature and matter power spectra. Our current results imply that the effects of the interaction and the dark energy equation of state are much stronger than the effect of the varying sound speed. This is expected since previous studies have shown that in order for the sound speed to leave an important observational imprint it has to be very small $c_{\rm s}^2=\mathcal{O}(10^{-3})$ \cite{Xia:2007km, dePutter:2010vy, Ballesteros:2010ks, Basse:2012wd}.

Moving forward, there is a vast parameter space to explore for these momentum transfer interactions: any coupling of the form $\beta_{n-2} Z^n$ where $n\neq 2$ would result in a non-unitary sound speed. One could also explore more complex forms for the coupling function. We intend to complete an MCMC analysis, and compare these variable sound speed models with a variety of other models, against current observational constraints.

In general, dark energy interactions result in modifications of the Euler and continuity equations. When the Euler equation is modified, as is the case for the models presented here, we have the breaking of the weak equivalence principle \cite{Koyama:2009gd}. The possibility of observing this effect with future surveys will be the subject of future work.

Recently, it has been argued in \cite{Heneka:2017ffk}  that ``cold dark energy'' with $c_{\rm s}^2=0$, which adds the clustering of the dark energy perturbations on top of the matter ones, is compatible with observations, and that future cluster growth data can help distinguish it from dark energy with sound speed one. In \cite{Heneka:2017ffk} the importance of having self consistent models in which both the equation of state and the sound speed of dark energy evolve with redshift was emphasised. Our formalism provides this and looking at the effects of our proposed models on cluster abundances will be the subject of future work.

\section{Acknowledgements} 
ML's research is supported by an STFC studentship. AP's work for this project was supported by a Dennis Sciama Fellowship at the University of Portsmouth. RC and RM are supported by the STFC grant ST/N000668/1. RM is also supported by the South Africa SKA Project. We would like to thank Ed Copeland, Antony Lewis, David Seery, and Thomas Tram for useful discussions.

\bibliographystyle{apsrev}
\bibliography{Variable_sound_speed_in_IDE_JCAP_revised}
\end{document}